\documentclass[twocolumn]{aastex63}
\usepackage{ifpdf}

\usepackage{color}

\shorttitle{Blue straggler stars in Collinder 261}
\shortauthors{Rain et al.}
\graphicspath{{./}{figures/}}

\begin{document}

\title{\large A study of the blue straggler population of the old open cluster Collinder 261}

\author[0000-0003-4009-8316]{Rain, M.~J.}
\affiliation{Dipartimento di Fisica e Astronomia, Universita’ di Padova, Vicolo Osservatorio 3, I-35122, Padova, Italy}

\author[0000-0002-0155-9434]{Carraro, G.}
\affiliation{Dipartimento di Fisica e Astronomia, Universita’ di Padova, Vicolo Osservatorio 3, I-35122, Padova, Italy}

\author[0000-0002-7091-5025]{Ahumada, J.~A.}
\affiliation{Observatorio Astron\'{o}mico, Universidad Nacional de C\'{o}rdoba, Laprida 854, 5000 C\'{o}rdoba, Argentina}

\author{Villanova, S.}
\affiliation{Departamento de Astronom\'{\i}a, Universidad de Concepci\'{o}n, 169 Casilla, Concepci\'{o}n, Chile}

\author{Boffin, H.}
\affiliation{ESO, Alonso de Cordova 3107, Santiago de Chile, Chile}
\affiliation{ESO, Karl-Schwarzschild Strasse 2, D-85748 Garching, Germany}

\author{Monaco, L.}
\affiliation{Departamento de Ciencias F\'{\i}sicas, Universidad Andr\'{e}s Bello, Rep\'{u}blica 220, 837-0134 Santiago, Chile}

\author{Beccari, G.}
\affiliation{ESO, Karl-Schwarzschild Strasse 2, D-85748 Garching, Germany}



\begin{abstract}

Blue Stragglers  are stars located in an unexpected region of the color-magnitude diagram of a stellar population,
as they appear bluer and more luminous than the stars in the turnoff region. They are ubiquitous, since they have been found among Milky Way field stars, in open and globular clusters, and also in other galaxies of the Local Group.
Here we present a study on the blue straggler population of the old and metal-rich open cluster 
\object[Collinder~261]{Collinder 261}, based on \emph{Gaia} DR2 data and on a multi-epoch radial velocity survey conducted with FLAMES@VLT. We also analyze the radial distribution of the blue straggler population to probe the dynamical status of the cluster.\\
Blue straggler candidates were identified first with \emph{Gaia} DR2, according to their position on the CMD, proper motions, and parallaxes. Their radial distribution was compared with  those of the main sequence, red giant, and red clump stars, to evaluate mass segregation. Additionally, their radial velocities (and the associated uncertainties) were compared with the mean radial velocity and the velocity dispersion of the cluster. When possible, close binaries and long-period binaries were also identified, based on the radial velocity variations for the different epochs. We also looked for yellow stragglers, i.e., possible evolved blue stragglers.\\
We found 53 blue stragglers members of Collinder~261,  six of them already identified in previous catalogs. Among the blue straggler candidates with radial velocity measurements, we found one long-period binary, five close-binary systems, three non-variable stars; we also identified one yellow straggler.

\end{abstract}

\keywords{star clusters and associations: general --- 
star clusters and associations: individual (Collinder 261) --- blue stragglers --- binaries: close}

\section{Introduction} 
\label{sec:intro}

\begin{figure*}
   \includegraphics[width=0.95\linewidth]{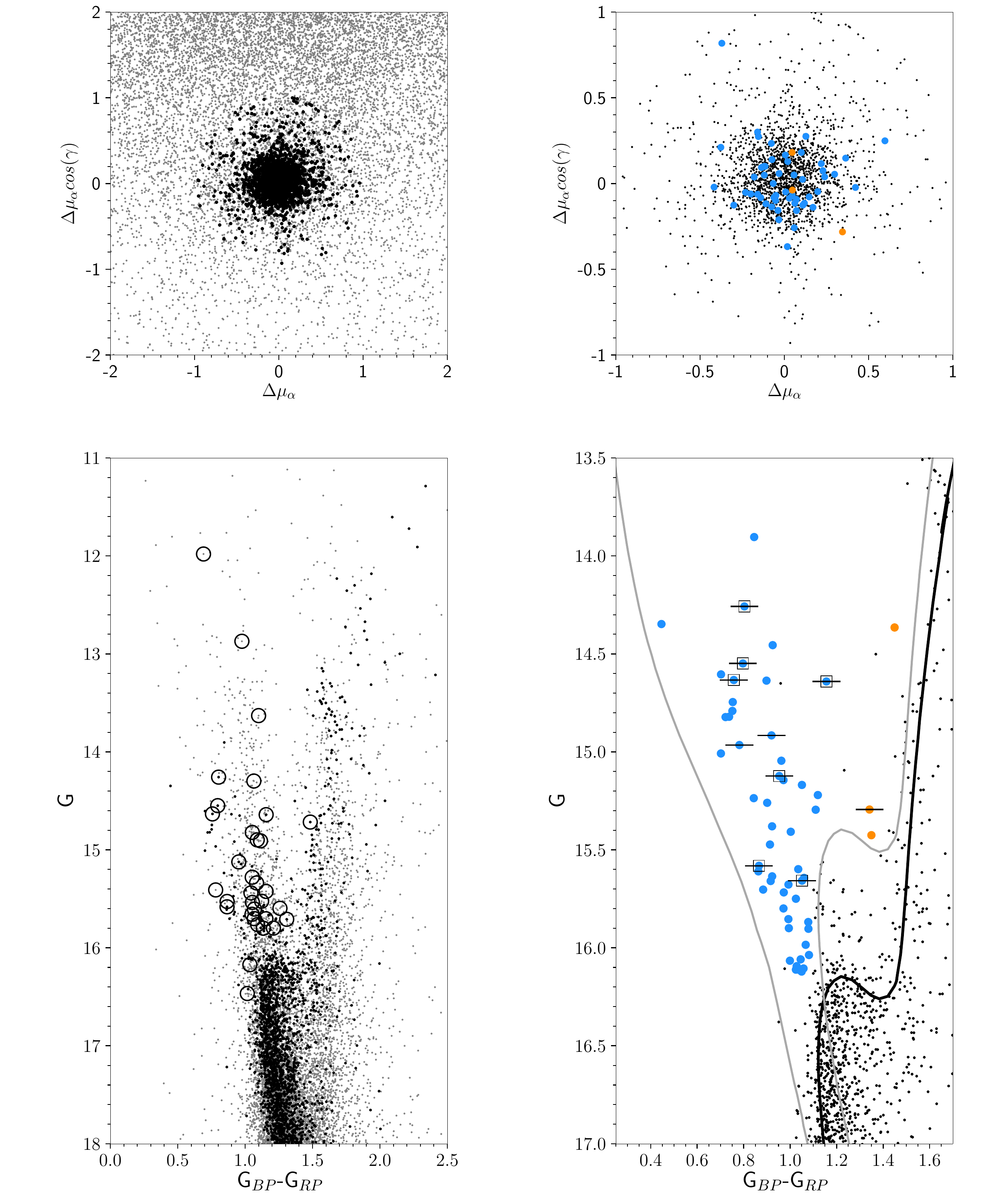}
    \caption{\textbf{Upper panels:} Proper motions plane of Collinder~261. In the left panel, gray dots represent all the sources listed by \emph{Gaia} DR2 within 15$'$ around the cluster center. Black filled circles are cluster members selected by \citet{Cantat-Gaudin_2018}.  In the right panel, we display a zoom of the center of the plane; here,  Collinder~261 members are the same as in the plot of the left, filled blue circles are BS candidates, and filled orange circles are yellow straggler candidates.  \textbf{Bottom panels:} Color-magnitude diagrams of Collinder~261. In the left panel, open black circles are BS candidates from \citet{Ahumada_2007}. The right panel displays a zoom into the region of BSs;  an isochrone from \citet{Bressan_2012}, shifted  the adopted cluster  age and metallicity, is shown in black solid line. The zero-age main sequence and the binary sequence are plotted in gray solid line. Black squares are BSs already identified by \citet{Ahumada_2007}. Stars with spectroscopic data are crossed with a horizontal line.}
    \label{fig:cmd+propermotions}
\end{figure*}

Blue Stragglers (BSs) are stars located in an unexpected region of the color-magnitude diagram (CMD) of a relatively old stellar population, since they appear bluer and more luminous than  stars in the turnoff (TO) region. Their existence is incompatible with the standard stellar evolution theory, which predicts that stars in this region of the CMD should have already left the main sequence because of their mass. Thus, these stars somehow managed  to gain mass and become a ``rejuvenated'' object. The formation mechanisms for the BSs are not yet fully understood; however, at present, there are two main leading scenarios: BSs could be the products of either direct stellar collisions \citep{Hills_1976}, or  mass-transfer activity in  close binary systems \citep{McCrea_1964}. Therefore, BSs can give information about the dynamical history of the cluster, the role of the dynamics on stellar evolution, the frequency of binary systems, and the contribution  of  binaries to cluster evolution. Hence, BSs certainly represent the link between standard stellar evolution  and cluster dynamics. Additionally, they are rather ubiquitous as they have been found in all kinds of stellar  environments: in the field, in open and globular clusters, and in galaxies of the Local Group.  An extensive review of their properties has been presented by \citet{Boffin_2015}. 

\begin{figure}
\centering
\includegraphics[width=\linewidth]{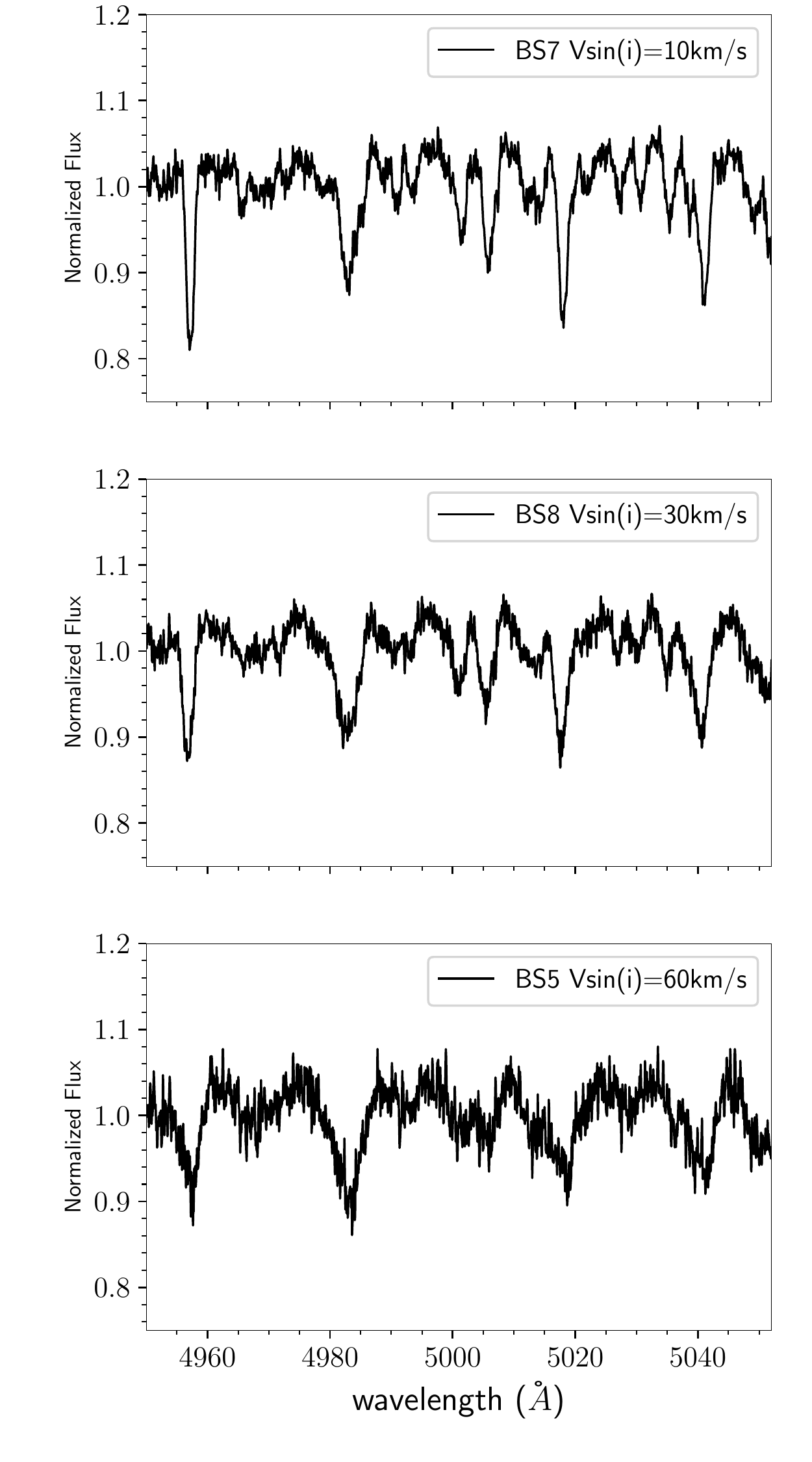}
\caption{Examples of observed spectra for three BSs with different rotational velocities, labeled on each panel, in the wavelength range between 4950 and 5050~\AA.}
\label{fig:wave-BS}
\end{figure}

Open clusters seem to be stellar systems where BSs find themselves particularly comfortable. The reason for this is not yet clear, and deserves more attention. The study of BS stars in open clusters is still limited to just a few cases, preventing their use as potential dynamical clocks, as it has been done in globular cluster (GC) environments \citep{Dalessandro_2014}. \citet[hereafter, F12]{Ferraro_2012} showed that GCs can be grouped into three different families based on the radial profile of their BS distributions. Clusters of \textit{Family~I}, or dynamically young GCs, show a flat distribution; in these systems the dynamical friction has not yet caused visible effects, even in the innermost regions. In \textit{Family~II} GCs, the  dynamical friction has become more efficient and the mass segregation has started, which has led to the presence of a peak at  the center, and a minimum at small  radii of the BS distribution. The outskirts of Family~II  clusters have still not been affected by the  dynamical friction, i.e., it has not reached the most remote BSs, and therefore there is a rise of the BS density in these outer regions. \citet{Bhattacharya_2019} recently studied the radial distribution of the very old open cluster  Berkeley~17 ($\sim$ 10 Gyr, \citealp{Kaluzny_1994}), and placed it in the Family~II class of GCs.  Finally, when the system is highly evolved, the external maximum disappears, and the only noticeable  peak in the distribution is the central one; GCs showing this profile are grouped in the \textit{Family~III}.\\
A few catalogs of BS stars in open clusters are available, but they are based on a purely photometric  selection \citep[hereafter d06 and AL07]{DeMarchi_2006, Ahumada_2007}. While useful, the photometric selection is not reliable enough  to allow the derivation of statistical properties of BS stars, since their membership is uncertain, and field stars tend to occupy the very same region of BS stars in the CMD \citep{Carraro_2008}. Only an accurate membership  assessment may let us know the real number of BS stars in a given cluster, and the evolutionary  status of each of them. So far, this has been done only for a handful of clusters. A very well studied cluster is M67, which harbors 24 BS stars (\citealp[and references therein]{Leonard_1996}). In this respect, however, perhaps the best studied case is NGC~188 \citep{Geller_2010, Geller_2011, Geller_2013}. These authors have found 21 bona~fide BS stars, with a binary fraction of 76\% $\pm$ 19\%. Among the 21 BSs, only four do nor show velocity variations, although  it cannot be discarded that they might be long-period variables ($P > 3000$~days), or that they are being seen pole-on.  Some binary BSs in open clusters belong to very long-period systems, and hence are difficult to detect spectroscopically and almost impossible to detect photometrically. The percentage of binaries among BS stars is significantly larger than in the cluster main sequence (MS), where it is of about 20\% \citep{Mathieu_2009,Geller_2012}. For all the BS binaries, the cited works derive period solutions. They find that the orbital period distribution of BS stars is quite different from that of MS binaries, with the majority of BS having orbital periods close to 1000 days, with most of them likely having a white dwarf companion. Recently, white dwarf companions have been detected for seven BSs in NGC~188 using far-ultraviolet HST observations \citep{Gosnell_2015}.  Similarly,  five of the six binary BSs of M67 have periods from 800 to 5000 days \citep{Latham_1996, Pribulla_2008}. Besides these two clusters, spectroscopic studies of BSs in open clusters are available only for individual stars, for example: NGC 6791, \citet{Brograard_2018}; NGC 6087, NGC 6087, NGC 6530, and Collinder 223, \citet{Aidelman_2018}; NGC 2141,  \citep{Luo_2015}.
 
 \begin{figure}
    \includegraphics[width=\linewidth]{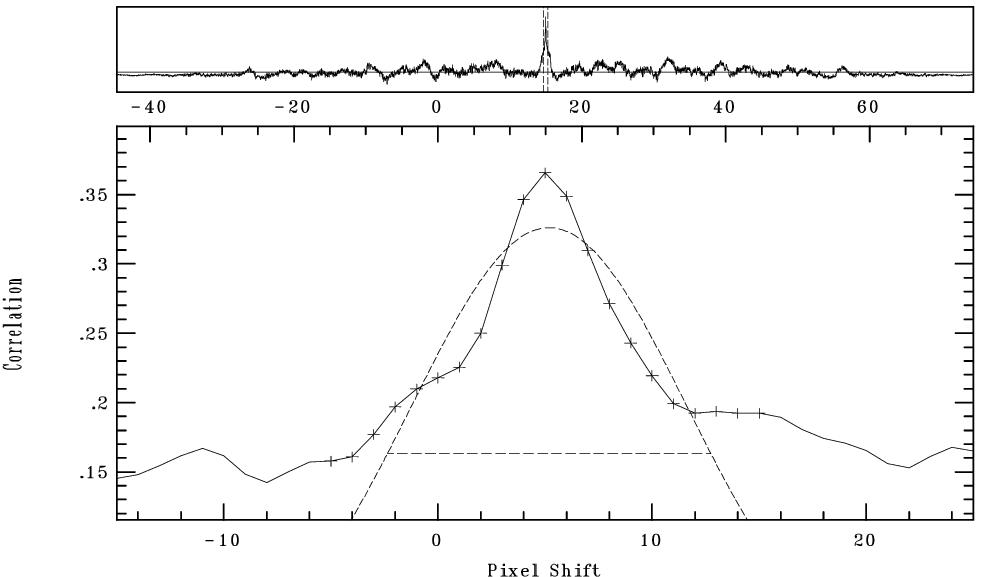}
    \includegraphics[width=\linewidth]{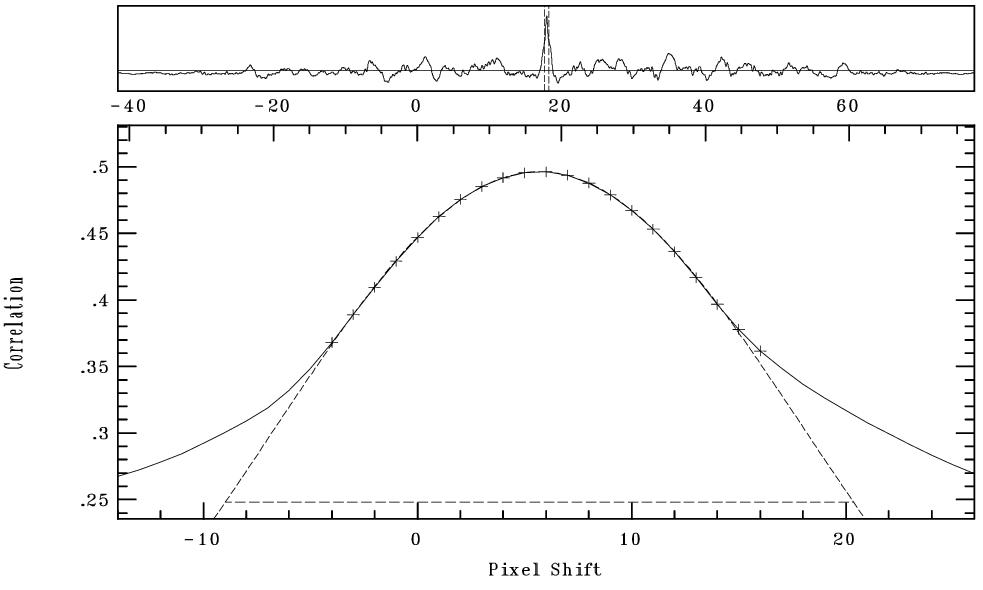}
    \caption{Cross-correlation peaks for star BS3, derived by the IRAF routine \texttt{fxcor}.  The top panel shows one of the computed synthetic templates  ($v\sin(i)$ equal to 10~km/s) cross-correlated with the spectrum of the star. Since the template is inadequate due to the low value of the rotational velocity, the cross-correlation curve is noisy. The bottom panel shows the same stellar spectrum cross-correlated with the correct template (with a rotational velocity of 30~km/s); in this case, the curve is smooth.}
    \label{fig:rot}
\end{figure}

Collinder 261,  or Harvard~6 (C1234$-$682, $\alpha=12^\mathrm{h}$37$^\mathrm{m}$57$^\mathrm{s}$,$\delta=-68$\arcdeg 22\arcmin 00$^{''}$, J2000.0), is one of the oldest open clusters of the Milky Way,  having an age from 7 to 9.3~Gyr \citep{Bragaglia_2006, Dias_2014}.  The cluster metallicity is close to solar, and reported values for the distance lie between 2.2 and 2.9~kpc, while its reddening  $E(B-V)$ has been estimated between  0.25 and 0.34~mag \citep{Mazur_1995,Gozzoli_1996, Bragaglia_2006, Dias_2014, Cantat-Gaudin_2018}. The  cluster parameters are summarized in Table~\ref{tab:Cr261_parameters}.
Due to the cluster location towards the galactic center, its color-magnitude diagram is  heavily contaminated by field stars. In this sense, the second and latest data release of \emph{Gaia} (DR2), which provides precise five-parameter astrometry (positions, parallaxes, and proper motions), as well as  three-band photometry ($G$, $G_\mathrm{BP}$, and  $G_\mathrm{RP}$ magnitudes) for more than one billion stars \citep{Lindegren_2018}, allows a proper study of Collinder~261 members and BS population with  great confidence. \citet{Gao_2018} and \citet{Cantat-Gaudin_2018} (hereafter CG18),  estimate $\sim$ 2000 members on the Collinder~261 area by applying the unsupervised clustering technique on the \emph{Gaia} data. \\
 The layout of the paper is as follows. In Section \ref{sec:datasets} we present the datasets used in this work. In section \ref{sec:phot_analysis} we describe the photometric analysis and the selection criteria of BS stars in open clusters, and the results of such selection 
for Collinder~261. Section \ref{sec:spec-analysis} explains how the spectra were reduced and the radial velocities 
extracted;  in this section we also define the criteria to establish membership and binary status of our targets. In Section \ref{sec:results-general} we discuss the results of the photometric and spectroscopic detection. Finally, in Section \ref{sec:conclusions}, after  a brief summary we give the conclusions of this study.

\begin{table*}
\centering
	\caption{Main parameters of Collinder~261.}
	\label{tab:Cr261_parameters}
	\begin{tabular}{c c c c c c c c c c c c} 
		\hline
  $l$$^{a}$ & $b$$^{a}$ & Distance$^{b}$ & $E(B-V)^{a}$ & $R_{c}^{1, c}$ & $R_{h}^{2, c}$ & $R_{a}^{1, a}$ &$\log(\mathrm{age})$$^{a}$ & [Fe/H] & RV$^{d}$ \\
 (deg) & (deg) & (kpc) & (mag) & (arcmin) & (arcmin) & (arcmin) & (yr) & (dex) & (km/s)  \\
\hline
 301.68 & $-0.53$ & 2.9 & 0.27 & 2.6 & 6.4 & 9.0 & 9.95 & $-0.01\pm0.11$ & $-25.44\pm0.93$  \\
		\hline
	\end{tabular}\\
	$^{1}$Core radius
	$^{2}$Half-mass radius
	$^{3}$Apparent radius\\
	$^{a}$\citet{Dias_2014}
	$^{b}$\citet{Cantat-Gaudin_2018}
	$^{c}$\citet{Vats_2017}
	$^{d}$\citet{Mishenina_2015}
\end{table*}

\section{Datasets} \label{sec:datasets}
\subsection{Photometric data} \label{subsec:photometric_data}

We used the Data Release~2 Archive\footnote{\url{https://gea.esac.esa.int/archive/}} of the 
European Space Agency mission \emph{Gaia} \citep{gaia16,gaia18}. For more than a billion stars, 
this survey provides a five-parameter astrometric solution: position, trigonometric parallax, and proper 
motion, as well as photometry in three broad-band filters ($G$, $G_\mathrm{BP}$, and $G_\mathrm{RP}$). The \emph{Gaia} catalog also gives radial velocities for about 7 million stars,  mostly brigther
than $G \sim 13$. The astrometric solution,  the photometric contents and validation, and
the properties and validation of radial velocities are described in
\citet{Lindegren_2018}, \citet{Evans_2018}, and \citet{Katz_2019}, respectively.

\subsection{Spectroscopic data} \label{subsec:spectroscopic_data}
Collinder~261 was observed with the fiber-fed spectrograph  FLAMES\footnote{\url{http://www.eso.org/sci/facilities/paranal/instruments/flames.html}}
at the Very Large Telescope (VLT) of the European Southern Observatory  (ESO, Paranal Observatory, Chile), using the combination of the mid-resolution spectrograph GIRAFFE and the  fiber link UVES.  The data were collected in two  periods:  October~2011 to March~2012, and October~2017 to April 2018, under ESO programs 088.D-0045(A) and 0100.D-0052(A).

The UVES fibers were allocated to the cluster's clump stars, whose membership is very solid, to set the zero point of the radial velocity. The reduction and  analysis of the UVES data are described by \citet{Mishenina_2015}. GIRAFFE was used with the setup HR8, which covers the wavelength range 491.7--516.3~nm, with a spectral resolution  $R \equiv \lambda/\Delta\lambda\equiv$~20,000. The integration time was $\sim$ 2,400~sec for all spectra. In total, Collinder~261 was observed in 4 epochs; some details of the observations are given in Table~\ref{tab:obs_details}. For the GIRAFFE data we just performed the sky-subtraction and normalization  using the IRAF\footnote{IRAF is distributed by the National Optical Astronomy Observatory, which is operated by the Association of Universities for Research in Astronomy, Inc.., under cooperative agreement with the National Science Foundation.} packages \texttt{sarith} and \texttt{continuum}, since the data were already reduced in Phase 3.

\begin{table}
	\centering
	\caption{Details of the spectroscopic observations carried out on February 24 and  March  1--6, 2012, and February 3--4, 2018, with a S/N between 15 and 120.}
	\label{tab:obs_details}
	\begin{tabular}{cccc} 
		\hline
		Epoch  &  MJD$_{\mathrm{start}}^{a}$  & MJD$_{\mathrm{end}}^{a}$ & Exposure \\
		 & (days) & (days) & (sec) \\
		\hline
	    1 & 55982.277794959 & 55982.3055727877  & 2400.0044\\
		2 & 55988.222295996 & 55988.2500738154	& 2400.0036\\
		3 & 55993.234082174 & 55993.2618600004  & 2400.0042\\
		4 & 58153.314142684 & 58153.3419204942  & 2400.0028\\
		\hline
	\end{tabular}
$^{a}$ Modified Julian date JD$-$24,00000.5)
\end{table}

\section{Photometric Analysis}\label{sec:phot_analysis}

For our photometry-based analysis, we took advantage of the selection of cluster members already performed by CG18, who used the membership assignment code UPMASK\footnote{\url{https://cran.r-project.org/web/packages/UPMASK/index.html}} (Unsupervised Photometric Membership Assignment in Stellar Clusters, \citealt{Krone_2014}). This procedure is based on proper motions and parallax selection criteria. They considered as cluster members those stars located over a radius twice as large as the value of $r_{\mathrm{DAML}} = 9'$ reported by \citet{Dias_2014}
(hereafter DAML02), with proper motions within 2~mas/yr, and with parallaxes within 0.5~mas from the cluster centroid. In this way, they selected about  2000  members of Collinder~261. For every star they estimated a membership probability from 0 to 100\%.  To calculate the cluster mean proper motion and parallax, they used only stars with probabilities above  50\%.

\citet{Cantat-Gaudin_2018} report the mean values  $\mu_{\alpha} \cos\delta = -6.35\pm 0.16$ (0.004) mas/yr, $\mu_{\alpha} = -2.70 \pm 0.16$ (0.004) mas/yr, and $\varpi = 0.315 \pm 0.082$ (0.002) mas.  Just to be cautious, we calculated our own proper motions values  following a similar procedure as that of \citet{Bhattacharya_2019}. We employed  the table access protocol (TAP) and the 
 astronomical data query language (ADQL), together with the Tool for OPerations on Catalogues And Table (TOPCAT),\footnote{\url{http://www.starlink.ac.uk/topcat/}}  to access the \emph{Gaia} data; for this,  
  we identified the Gaia DR2 counterparts of confirmed Collinder~261 members, as follows.
 \citet{deSilva_2007} measured radial velocities, metallicities, and chemical abundances (Mg, Si, Ca, Mn, Fe, Ni, Zr, and Ba) of 12 giant stars, and confirmed their membership; we cross-correlated the position on the sky of these stars and the \emph{Gaia} DR2 catalog, looking for  the nearest neighbors within 1$''$. Our mean proper motions  and parallax values  are: $\mu_{\alpha}\cos\delta = -6.35 \pm 0.13$ mas/yr, $\mu_{\alpha} = -2.73 \pm 0.14$ mas/yr, and $\varpi = 0.321\pm0.019$ mas, which are in complete agreement with those of CG18 and \citet{Gao_2018} ($\mu_{\alpha} \cos\delta = -6.340 \pm 0.006$ mas/yr, $\mu_{\alpha} = -2.710 \pm 0.004$ mas/yr,  and  $\varpi = 0.3569\pm 0.0027$ mas). On the other hand, even considering the errors, our results are far from the values given by  DAML02, namely,  $\mu_{\alpha}\cos\delta = -0.65 \pm4.94$ mas/yr, and  $\mu_{\alpha} = -0.51 \pm 3.76$ mas/yr.
 
 \citet{Arenou_2018}  report the differences between the average zero points of proper motions and parallaxes from DAML02 and \emph{Gaia} catalogs. They found $\mu_{Gaia}-\mu_{\rm DAML02}=0.0 \pm 0.19$ mas/yr and $0.41\pm 0.18$ mas/yr  for  $\mu_{\alpha}$ and $\mu_{\alpha}\cos\delta$  respectively, and  $\varpi_{Gaia}-\varpi_{\rm DAML02}=-0.064\pm 0.17$ mas for the parallaxes. \citet{Arenou_2018} do not find evidence for the presence of significant systematic errors in the \emph{Gaia}~DR2 proper motions. Therefore, the differences between our results and the published values of DAML02 seem unlikely to be caused by systematic errors in the \emph{Gaia} data. Since  our values were calculated using giant stars of confirmed membership, we suggest that the DAML02 proper motions might suffer from the lack of reliable cluster membership, as well as from  significant contamination by field stars; these may be the main reasons of the discrepancies noted above.

\subsection{Identification of the stragglers} \label{subssec:identification_BSS}

The region populated by the BSs in a CMD is well known today. According to \citet{Ahumada_1995, Ahumada_2007} 
this area is delimited, to the left, by the zero-age main sequence (ZAMS); to the right, by the turnoff color; 
and down, by the magnitude at which the observed sequence of the cluster separates from the ZAMS.  After applying a membership selection criterion, all stars falling   in this region can be considered as genuine  blue stragglers with good certainty. Following AL07, we superimposed an 8.7~Gyr isochrone and a ZAMS of solar metallicity from  \citet{Bressan_2012}, using $(m-M)_{0} = 11.86$ and $E(B-V) = 0.27$, to the observed CMD (Figure \ref{fig:cmd+propermotions}). To  not include spurious stars,  we further constrained this region by plotting the equal-mass binary locus (dashed line)  obtained by shifting the isochrone by 0.753 in  $G$  toward  brighter magnitudes; in this way, we expect that   binaries containing normal main-sequence TO stars are excluded. These stars may appear as  stragglers, but their components may not be such---see, e.g., \citet{Hurley_1998} for a discussion of this sequence.

\citet{Ahumada_2007} estimated a red limit of $(B-V)\simeq 0.86$ for the BS area of Collinder~261. 
To impose the same limit in the \emph{Gaia} system, we used the relation of \citet{Jordi_2010}:
\begin{equation}
C_{1} = 0.0187 + 1.6814\, C_{2} - 0.3357\, C_{2}^{2} + 0.117\, C_{2}^{3}.
\end{equation} 
With $C_{2} = 0.86$, 
 it results $C_{1} = (G_\mathrm{BP}-G_\mathrm{RP})   = 1.20$. 
Stars redder than this limit can be considered as possible yellow stragglers.

Finally, we adopted an upper limit of 2.5 magnitudes above the TO for massive BSs.
 
In the top panel of Fig. \ref{fig:cmd+propermotions} we plotted the distribution of proper motions for stars in the Collinder~261 area. Gray dots  represent \emph{Gaia} DR2 stars within $15'$ of the cluster center. Black filled circles are the cluster members of CG18. Light blue filled circles are the BS candidates, and orange filled circles are the yellow stragglers  candidates. In the bottom panel of Fig. \ref{fig:cmd+propermotions} we plotted the color-magnitude diagram of Collinder~261. The symbols are the same of the top panel. Open black circles are the BS candidates  identified by AL07. Only 6 of 54 are classified as members according to \emph{Gaia} data. The BS sample of AL07 follows the galactic field trend and is very different from the BS  population found with \emph{Gaia}. In total we identified 55 BS and five YS candidates.

Most of the BSs are within 0.5 mas/yr from the cluster mean, while only four candidates 
lie outside this region. The membership probabilities of the  latter are: 10\%, 30\%, 50\%, and 70\%. The remaining candidates have probabilities above  70\%, so we decided to leave out all the stars with  probability below 50\% to define a bona~fide, non-spurious BS population. The same criterion was applied to the yellow straggler candidates, from the five candidates, two stars lie outside the region mentioned above; they have membership probabilities of 30\% and 40\%, and were left out of our sample.

Summing up the results of this Section, Collinder~261 has 53 BS and three YS candidates. These are listed in Table \ref{tab:BS_phot}. When available, additional classification according to their binary nature, as reported in the astronomical database SIMBAD \citep{Wenger_2000}, is also listed. The eclipsing, close binaries found among the BSs of Collinder~261 are of the following types: $\beta$~Lyrae, Algol, and W~Ursae~Majoris (W~UMa). Binaries classified as $\beta$~Lyrae  are semidetached systems, i.e., one of the components of the pair is filling its critical Roche lobe, the stars are close enough that they are gravitationally distorted, and the  periods are usually longer than 1 day. Algol variables are also semidetached binaries, whose components  have  spherical or slightly ellipsoidal shapes.  These stars have an extremely wide range of observed periods, from 0.2 to over 10,000 days. In the W~UMa-type stars the components are in contact or almost in contact, and share a common envelope of material; the orbital period can  be  of just a few hours.

\section{Spectroscopic Analysis} \label{sec:spec-analysis}

This is the first high-resolution spectroscopic analysis of the BS population in Collinder~261.  Unfortunately, not all the candidates were observed with FLAMES since, when we were allocated the observational time, we used the BS list of AL07 to select the targets, a list very different from that found in this work using \emph{Gaia}.
The spectroscopic analysis was carried out on nine out of the 53 BS in our  list, and on one  of the three YS candidates identified.

\subsection{Radial Velocities} 
\label{subsec:rv}

For each spectrum, radial velocities were determined with the IRAF \texttt{fxcor}  cross-correlation task  \citep{Tonry_Davis_1979}. Stellar spectra were cross-correlated with synthetic templates obtained with the SPECTRUM code\footnote{\url{http://www.appstate.edu/~grayro/spectrum/spectrum.html}} 
\citep{Gray_Corbally_1994}.  Each synthetic spectrum was computed adopting the atomic and 
molecular data file \texttt{stdatom.dat}, which includes solar atomic abundances from 
\citet{Grevesse_1998}, and the linelist \texttt{luke.lst}, suitable for mid-B- to K-type stars.  Model atmospheres were calculated with the code ATLAS9 \citep{Castelli_Kurucz_2003}.

In Figure \ref{fig:rot} we plotted, as an example,  the spectra of three of our targets. The selection of a proper template for each star was mandatory because the targets have different rotational velocities (see \citealt{Mucciarelli_2014} for more information). We computed a set of templates with different rotational velocities $v\sin(i)$ ranging from 10 to 100~km/s, adopting the values: for the effective temperature,   $T_{\mathrm{eff}} = 6000$~K, for the surface gravity, $\log g = 4.5$~cm/s$^2$, and for the micro-turbulence, $\xi = 0.0$~km/s.  We carefully compared the spectrum of each star  with the templates, and we visually estimated  the rotation rate from the profiles of the spectral lines. However, if the template had too low a rotational velocity, the shape turned out to be very noisy because the profiles of the lines and the spectrum noise were mapped together. In these cases, we had to increase the rotational velocity of the template for the cross-correlation procedure. The derived rates are considered as upper limits. An example of this procedure is illustrated in Figure \ref{fig:rot}.

The radial velocities measurements for the blue and yellow stragglers are reported in Table \ref{tab:rv_blue}.

\startlongtable
\begin{deluxetable*}{cccccccccc}\label{tab:BS_phot}
\centering
\tablecaption{Blue and yellow straggler candidates from Gaia DR2 data}
\startdata
\emph{Gaia} DR2 Source Id.  & $G$ & $G_\mathrm{BP}$ & $G_\mathrm{RP}$  & $\mu_{\alpha}cos(\delta)$ & 
 $\mu_{\alpha}$   & Parallax $\bar{\omega}$ & $P_{\mathrm{Memb}}$ & Class$^{a}$ \\
  & (mag) & (mag) & (mag) & (mas/yr) & (mas/yr) & (mas) & (\%)& \\
 \hline
 5856432828647768960 & 15.00 & 15.28  & 14.57 & $-$6.485 $\pm$ 0.038 & $-$2.636 $\pm$ 0.043 & 0.304 $\pm$ 0.024 & 100 & BS \\
 5856437089255328384 & 15.23 & 15.57  & 14.73 & $-$6.302 $\pm$ 0.050 & $-$2.800 $\pm$ 0.044 & 0.372 $\pm$ 0.029 & 100 & BS\\
 5856511271928032256 & 14.74 & 15.03  & 14.28 & $-$6.399 $\pm$ 0.037 & $-$2.799 $\pm$ 0.041 & 0.318 $\pm$ 0.028 & 100 & BS\\ 
 5856530272864404224 & 15.40 & 15.80  & 14.80 & $-$6.291 $\pm$ 0.052 & $-$2.988 $\pm$ 0.052 & 0.306 $\pm$ 0.036 & 100 & BS\\ 
 5856527455365735680 & 15.26 & 15.61  & 14.71 & $-$6.278 $\pm$ 0.053 & $-$2.821 $\pm$ 0.056 & 0.290 $\pm$ 0.037 & 100 & BS\\ 
 5856516597687652352 & 15.89 & 16.31  & 15.31 & $-$6.424 $\pm$ 0.057 & $-$2.865 $\pm$ 0.068 & 0.295 $\pm$ 0.042 &  90 & BS\\
 5856527455365740032 & 15.16 & 15.59  & 14.54 & $-$6.464 $\pm$ 0.042 & $-$2.630 $\pm$ 0.042 & 0.337 $\pm$ 0.028 &  90 & BS\\ 
 5856527386646253312 & 14.54 & 14.85  & 14.05 & $-$6.383 $\pm$ 0.034 & $-$2.940 $\pm$ 0.036 & 0.297 $\pm$ 0.024 &  90 & BS\\ 
 5856527386646253440 & 15.12 & 15.45  & 14.50 & $-$6.423 $\pm$ 0.043 & $-$2.589 $\pm$ 0.055 & 0.333 $\pm$ 0.032 & 100 & BS\\
 5856527524085206656 & 13.90 & 14.24  & 13.39 & $-$6.319 $\pm$ 0.030 & $-$2.815 $\pm$ 0.030 & 0.297 $\pm$ 0.020 & 100 & BS\\ 
 5856517010004497792 & 15.65 & 16.03  & 15.11 & $-$6.649 $\pm$ 0.053 & $-$2.855 $\pm$ 0.060 & 0.340 $\pm$ 0.036 &  90 & BS, EB $\beta$ Lyrae\\ 
 5856528348719355648 & 14.63 & 14.92  & 14.16 & $-$6.388 $\pm$ 0.038 & $-$2.886 $\pm$ 0.039 & 0.351 $\pm$ 0.026 & 100 & BS\\  
 5856528623597294720 & 14.82 & 15.08  & 14.34 & $-$6.766 $\pm$ 0.044 & $-$2.751 $\pm$ 0.043 & 0.334 $\pm$ 0.032 &  70 & BS\\  
 5856517078723954944 & 15.29 & 15.68  & 14.57 & $-$6.489 $\pm$ 0.046 & $-$2.813 $\pm$ 0.049 & 0.322 $\pm$ 0.032 & 100 & BS\\  
 5856527558444960128 & 14.60 & 14.86  & 14.16 & $-$6.110 $\pm$ 0.037 & $-$2.691 $\pm$ 0.041 & 0.377 $\pm$ 0.026 & 100 & BS\\ 
 5856527455365743488 & 14.82 & 15.09  & 14.37 & $-$6.427 $\pm$ 0.044 & $-$2.495 $\pm$ 0.050 & 0.277 $\pm$ 0.033 & 100 & BS\\  
 5856527661524172800 & 14.91 & 15.28  & 14.36 & $-$6.343 $\pm$ 0.041 & $-$2.564 $\pm$ 0.045 & 0.283 $\pm$ 0.028 & 100 & BS, EB $\beta$ Lyrae\\
 5856527730243676800 & 15.60 & 15.94  & 15.07 & $-$6.330 $\pm$ 0.055 & $-$2.601 $\pm$ 0.051 & 0.331 $\pm$ 0.036 & 100 & BS\\ 
 5856528383078700288 & 15.71 & 16.11  & 15.14 & $-$6.381 $\pm$ 0.059 & $-$2.671 $\pm$ 0.063 & 0.281 $\pm$ 0.039 & 100 & BS, EB Algol\\
 5856528726676132096 & 15.85 & 16.23  & 15.24 & $-$6.242 $\pm$ 0.060 & $-$2.705 $\pm$ 0.063 & 0.345 $\pm$ 0.042 & 100 & BS\\
 5856528520517692032 & 15.64 & 16.06  & 15.00 & $-$6.548 $\pm$ 0.059 & $-$2.790 $\pm$ 0.053 & 0.320 $\pm$ 0.037 & 100 & BS\\  
 5856528589237173120 & 15.22 & 15.65  & 14.53 & $-$6.222 $\pm$ 0.046 & $-$2.455 $\pm$ 0.047 & 0.374 $\pm$ 0.032 & 100 & BS\\
 5856529242072194816 & 15.65 & 16.09  & 15.04 & $-$6.402 $\pm$ 0.063 & $-$2.829 $\pm$ 0.060 & 0.310 $\pm$ 0.041 & 100 & BS\\  
 5856517284882496128 & 14.63 & 15.00  & 14.10 & $-$6.411 $\pm$ 0.037 & $-$2.811 $\pm$ 0.038 & 0.268 $\pm$ 0.025 & 100 & BS\\
 5856528348719366528 & 15.47 & 15.83  & 14.91 & $-$5.985 $\pm$ 0.052 & $-$2.582 $\pm$ 0.054 & 0.344 $\pm$ 0.034 &  90 & BS\\ 
 5856515601255190272 & 14.25 & 14.57  & 13.76 & $-$6.293 $\pm$ 0.031 & $-$2.679 $\pm$ 0.033 & 0.313 $\pm$ 0.023 & 100 & BS, EB Algol\\ 
 5856514879700797696 & 14.34 & 14.49  & 14.05 & $-$6.415 $\pm$ 0.032 & $-$2.729 $\pm$ 0.035 & 0.322 $\pm$ 0.021 & 100 & BS\\ 
 5856528550550957184 & 14.96 & 15.25  & 14.47 & $-$6.278 $\pm$ 0.043 & $-$2.888 $\pm$ 0.042 & 0.281 $\pm$ 0.030 &  90 & BS, EB Algol\\
 5856529379511194240 & 14.79 & 15.07  & 14.32 & $-$6.120 $\pm$ 0.038 & $-$2.655 $\pm$ 0.039 & 0.297 $\pm$ 0.026 & 100 & BS\\
 5856527524085202048 & 15.58 & 15.90  & 15.04 & $-$6.579 $\pm$ 0.051 & $-$2.782 $\pm$ 0.051 & 0.307 $\pm$ 0.034 & 100 & BS\\ 
 5856432656849042688 & 16.03 & 16.50  & 15.41 & $-$6.529 $\pm$ 0.068 & $-$2.692 $\pm$ 0.060 & 0.290 $\pm$ 0.037 & 100 & BS\\
 5856420729693697408 & 16.12 & 16.57  & 15.52 & $-$6.473 $\pm$ 0.064 & $-$2.631 $\pm$ 0.064 & 0.334 $\pm$ 0.040 & 100 & BS\\
 5856419531427900672 & 15.90 & 16.36  & 15.28 & $-$6.469 $\pm$ 0.059 & $-$2.681 $\pm$ 0.059 & 0.304 $\pm$ 0.040 &  80 & BS\\
 5856437742090215680 & 15.79 & 16.20  & 15.22 & $-$6.727 $\pm$ 0.069 & $-$2.519 $\pm$ 0.055 & 0.245 $\pm$ 0.038 &  80 & BS\\
 5856436298981175168 & 16.09 & 16.52  & 15.49 & $-$6.504 $\pm$ 0.064 & $-$2.456 $\pm$ 0.057 & 0.343 $\pm$ 0.037 & 100 & BS\\
 5856436367700747392 & 15.98 & 16.43  & 15.37 & $-$6.249 $\pm$ 0.073 & $-$2.549 $\pm$ 0.066 & 0.436 $\pm$ 0.039 &  80 & BS\\
 5856483303104274560 & 15.12 & 15.54  & 14.57 & $-$5.753 $\pm$ 0.047 & $-$2.481 $\pm$ 0.040 & 0.403 $\pm$ 0.026 &  50 & BS\\
 5856512680677308160 & 16.10 & 16.55  & 15.49 & $-$6.342 $\pm$ 0.065 & $-$2.779 $\pm$ 0.270 & 0.270 $\pm$ 0.050 & 100 & BS\\
 5856528962867671936 & 16.06 & 16.46  & 15.46 & $-$6.509 $\pm$ 0.075 & $-$2.793 $\pm$ 0.069 & 0.204 $\pm$ 0.047 & 100 & BS\\
 5856528486157936384 & 16.05 & 16.47  & 15.42 & $-$6.131 $\pm$ 0.071 & $-$2.614 $\pm$ 0.064 & 0.412 $\pm$ 0.045 & 100 & BS\\
 5856528486157941376 & 16.11 & 16.52  & 15.50 & $-$6.240 $\pm$ 0.074 & $-$2.866 $\pm$ 0.066 & 0.320 $\pm$ 0.044 & 100 & BS\\
 5856527764603421184 & 15.86 & 16.31  & 15.23 & $-$6.152 $\pm$ 0.073 & $-$2.777 $\pm$ 0.062 & 0.278 $\pm$ 0.041 & 100 & BS\\
 5856529001554008064 & 14.64 & 15.12  & 13.97 & $-$6.288 $\pm$ 0.037 & $-$2.845 $\pm$ 0.040 & 0.306 $\pm$ 0.026 & 100 & BS\\
 5856528761035882496 & 15.38 & 15.74  & 14.82 & $-$6.203 $\pm$ 0.053 & $-$2.808 $\pm$ 0.053 & 0.226 $\pm$ 0.035 & 100 & BS, EB W UMa\\
 5856529276431965696 & 16.37 & 16.76  & 15.81 & $-$6.194 $\pm$ 0.077 & $-$2.974 $\pm$ 0.077 & 0.304 $\pm$ 0.053 & 100 & BS\\
 5856517211839051392 & 15.59 & 16.00  & 14.96 & $-$6.721 $\pm$ 0.063 & $-$1.912 $\pm$ 0.066 & 0.488 $\pm$ 0.043 &  70 & BS\\
 5856527558444963200 & 15.67 & 16.06  & 15.07 & $-$6.231 $\pm$ 0.057 & $-$2.845 $\pm$ 0.069 & 0.236 $\pm$ 0.038 & 100 & BS\\
 5856529276431950080 & 15.04 & 15.43  & 14.47 & $-$6.460 $\pm$ 0.046 & $-$2.848 $\pm$ 0.047 & 0.259 $\pm$ 0.031 & 100 & BS\\
 5856527558444950656 & 15.74 & 16.10  & 15.08 & $-$6.052 $\pm$ 0.093 & $-$2.676 $\pm$ 0.093 & 0.178 $\pm$ 0.062 & 100 & BS\\
 5856533326555127168 & 16.11 & 16.52  & 15.48 & $-$6.332 $\pm$ 0.070 & $-$3.098 $\pm$ 0.068 & 0.254 $\pm$ 0.047 & 100 & BS\\
 5856515665648722688 & 15.63 & 15.99  & 15.07 & $-$6.180 $\pm$ 0.056 & $-$2.869 $\pm$ 0.054 & 0.216 $\pm$ 0.038 & 100 & BS, EB Algol\\
 5856515669974676480 & 15.70 & 16.01  & 15.13 & $-$5.928 $\pm$ 0.081 & $-$2.753 $\pm$ 0.074 & 0.440 $\pm$ 0.056 &  90 & BS\\
 5856519037229366144 & 14.45 & 14.83  & 13.90 & $-$5.575 $\pm$ 0.030 & $-$2.430 $\pm$ 0.031 & 0.530 $\pm$ 0.021 &  80 & BS\\
 5856435130750056576 & 14.36 & 15.02  & 13.57 & $-$6.005 $\pm$ 0.03  & $-$3.012 $\pm$ 0.029  & 0.145 $\pm$  0.018 &  60  & YS \\
 5856527622837778176 & 15.29 & 15.89  & 14.55 & $-$6.302 $\pm$ 0.05  & $-$2.768 $\pm$ 0.049  & 0.304 $\pm$  0.033 & 100  & YS \\
 5856515601255187712 & 15.42 & 16.02  & 14.67 & $-$6.304 $\pm$ 0.04  & $-$2.549 $\pm$ 0.048  & 0.278 $\pm$  0.033 & 100  & YS \\
 \hline
\enddata
\vspace*{+0.25cm}$^{a}$ Classification of the stars according to their binary nature, as reported in the astronomical database SIMBAD \citep{Wenger_2000}
 \end{deluxetable*}

\begin{table*}
\centering
 \caption{Individual radial velocities measurements for blue and yellow stragglers in our sample. Binary classification according to their radial velocity variability and rotational rates are reported in the last two columns.}
 \label{tab:rv_blue}
 \begin{tabular}{l c c c c c c c} 
 \hline 
 ID$^{a}$ & \emph{Gaia} DR2 Source Id. & RV$_{1}$ & RV$_{2}$ & RV$_{3}$ & RV$_{4}$ &  Classification$^{a}$ & $< v\sin(i)$ \\
 &  & (km/s) & (km/s) & (km/s) & (km/s)& &(km/s)\\
 \hline
 BS1  & 5856527386646253312 & $-29.61\pm3.69$   & $-17.00\pm3.39$  & $+01.56\pm2.61$    & $-20.62\pm4.61$ &  M, CB & 30 \\
 BS2  & 5856528348719355648 & $-26.09\pm1.55$   & $-26.21\pm2.59$  & $-24.39\pm1.37$    & $-25.46\pm1.41$ &  M & 10 \\
 BS3  & 5856527524085202048 & $-24.23\pm5.45$   & $-17.25\pm7.11$  & $-00.02\pm3.22$    & $-02.55\pm5.79$ &  M, CB & 30 \\
 BS4  & 5856528550550957184 & $-35.30\pm8.45$   & $-24.03\pm11.07$ & $-1.80\pm6.59$     & $-15.37\pm6.72$ &  M, CB & 40\\
 BS5  & 5856527661524172800 & $+12.44\pm14.02$  & $-05.36\pm8.46$  & $-38.77\pm19.04$   & $+19.70\pm11.51$&  M, CB & 60  \\
 BS6  & 5856527386646253440 & $-27.45\pm5.14$   & $-29.75\pm3.30$  & $-16.24\pm3.31$    & $-12.90\pm3.35$ &  M, LP & 30\\
 BS7  & 5856529242072194816 & $-24.08\pm0.40$   & $-24.30\pm0.40$  & $-24.02\pm0.57$    & $-23.91\pm0.43$ &  M & 10\\
 BS8  & 5856515601255190272 & $-35.96\pm4.60$   & $+10.20\pm3.97$  & $-63.48\pm5.40$    & $-66.91\pm4.47$ &  M, CB & 30 \\
 BS9  & 5856529001554008064 & $-24.14\pm0.76$   & $-25.11\pm0.86$  & $-23.84\pm1.11$    & $-22.82\pm0.74$ &  M & 10\\
 \hline
 YS1  & 5856527622837778176 & $-24.51\pm0.34$ & $-24.65\pm0.34$ & $-24.56\pm0.30$ & $-24.03\pm0.45$ &  M & 10\\
 \hline 
 \end{tabular} \\
 $^a$ This work.
  \end{table*}

\subsection{Errors}
\label{subsec:errors}
\begin{figure}
\includegraphics[width=\linewidth]{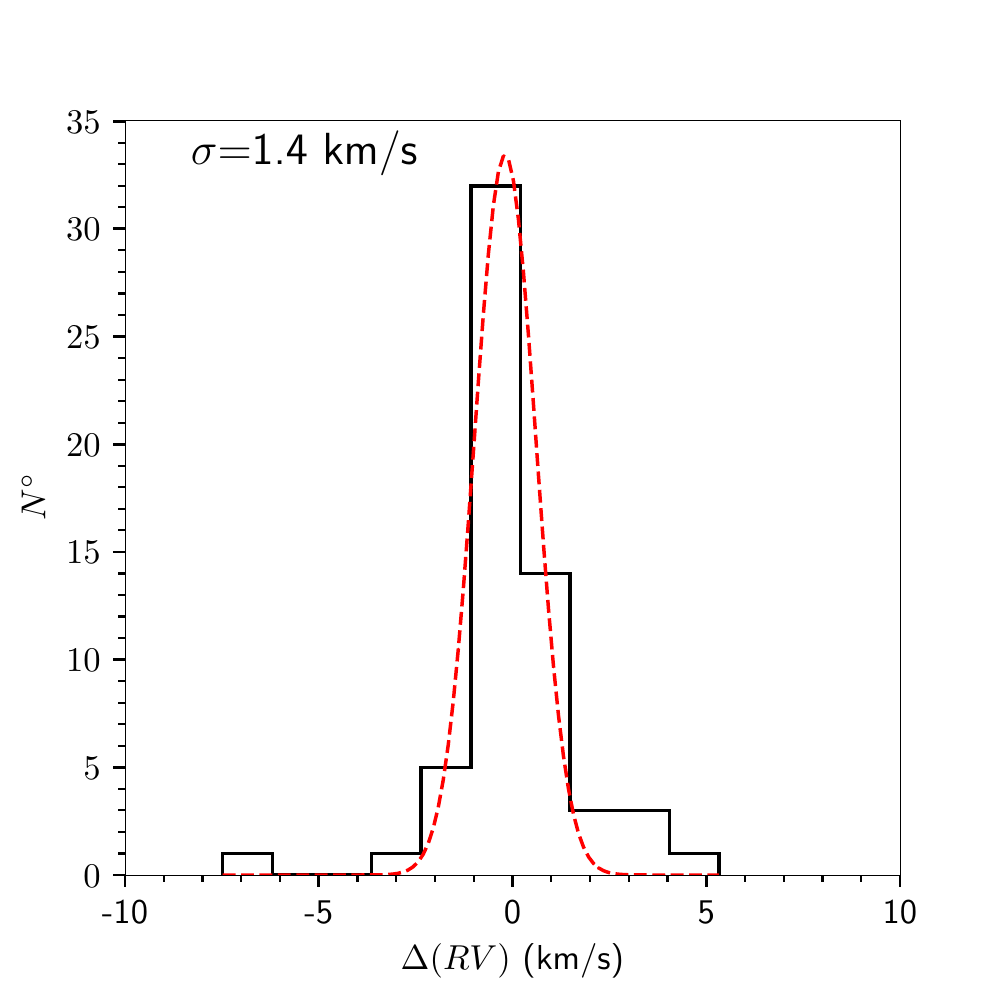}
\includegraphics[width=\linewidth]{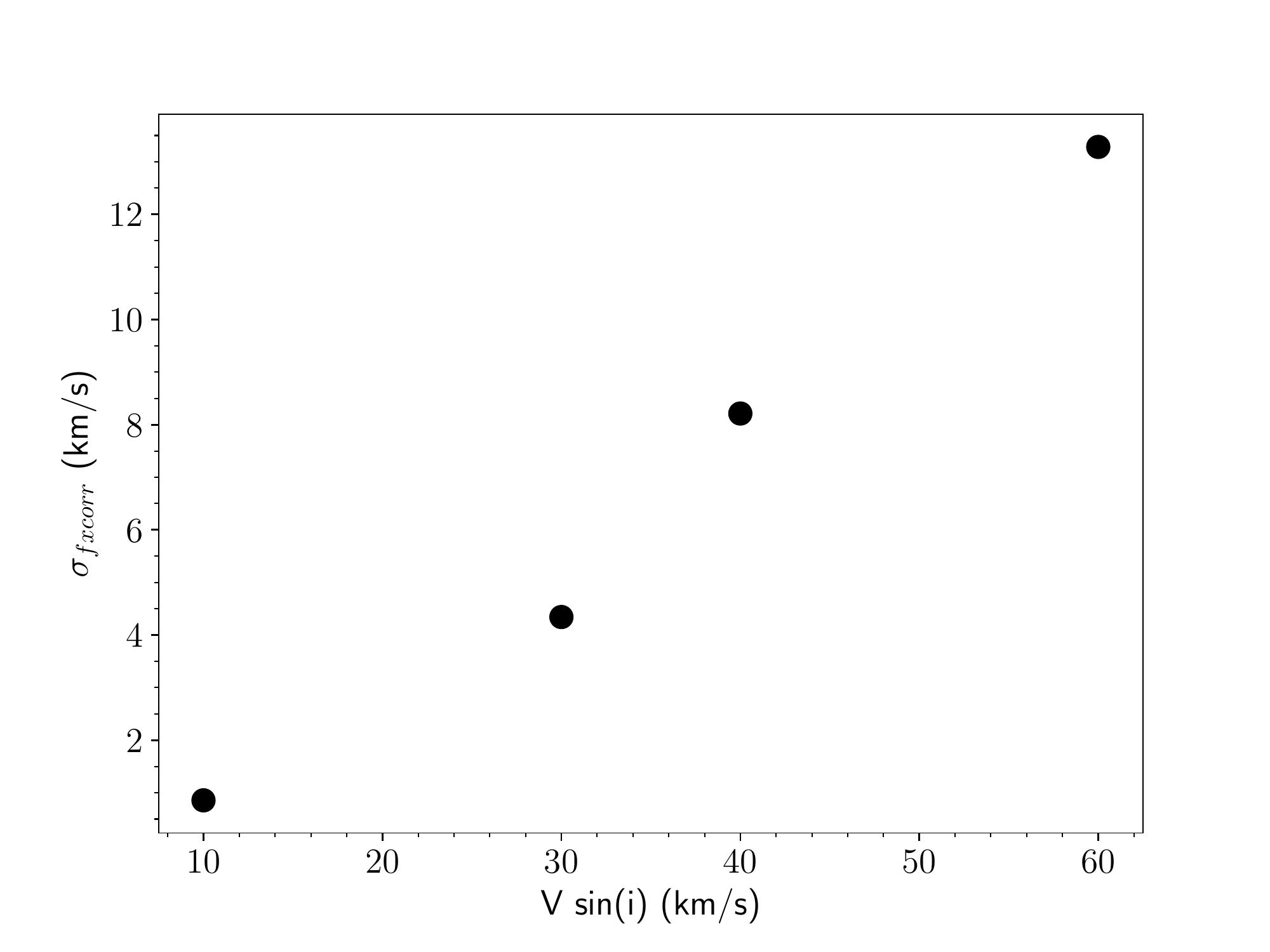}
\caption{\textbf{Upper panel:} Histogram of the differences---divided by the root square of 2---between pairs of radial velocities measurements for the same star.  The best-fitting Gaussian to the distribution is plotted, and its standard deviation $\sigma$  is  indicated. \textbf{Bottom panel:} \texttt{fxcor} mean errors as a function of the upper rotational rates estimated in this work. See text for details.}
    \label{fig:err_rv}
\end{figure}

We consider the errors returned by \texttt{fxcor}  as conservative estimates of the true uncertainties of the radial velocity. For each star we have four radial velocity measurements and \texttt{fxcor}  error estimations. We computed the \texttt{fxcor} error for each star, and for each pair of measurements we calculated the radial velocity difference divided by the root square of 2. We then built  the distribution histogram and fitted a Gaussian. We take the standard dispersion $\sigma$ of the Gaussian as the  true radial velocity error. We plotted the histograms together with the Gaussian fit and the true error in the upper panel of Figure \ref{fig:err_rv}. Additionally, we calculated the mean \texttt{fxcor} error for each rotational rate (estimated as we described above in Subsection \ref{subsec:rv}). Our results are plotted in the bottom panel of Figure \ref{fig:err_rv}. The typical uncertainties for the slow rotators stars ($v \sin(i)=10$~km/s) is about 1~km/s. Stars rotating with velocities ranging approximately between 30 and 60~km/s have errors from  4 to 14~km/s. Similar uncertainties values were found by \citet{Mucciarelli_2014} on their BS sample. Therefore, we decided to adopt the \texttt{fxcor} error as a conservative estimation for the radial velocity uncertainty.

\subsection{Membership and evolutionary status}\label{subsec:Member_evol_status}

Comparing our  radial velocities with the mean value derived by \citet{Mishenina_2015} for the clump stars,  we can now try to determine the  membership status of the objects in our sample. In what follows we
will assume that BSs are  the result of collisions, or that they are  binary systems, with either relatively short periods (a few days or less),  or rather long ones (about 1000 days). Tree epochs of observation are separated by a few days ($\sim$6), and the fourth epoch is about 6 years later. To assess membership,  the radial velocities of the stars can be compared to the mean radial velocity of the cluster, taking into account the error bars---as derived in Section~\ref{subsec:rv}---and the possibility of binarity.

The TO mass of Collinder~261 is about 1.1~$M_{\odot}$ \citep{Bragaglia_2006};  if the companions are main-sequence stars, they have to be less massive since, by definition, they are the secondary components. If the systems are post-mass-transfer ones containing a white dwarf (e.g., \citealt{Gosnell_2014}), then the mass of the companions are more likely peaked around 0.6~$M_{\odot}$, as found in NGC~188 \citep{Geller_2011}. We can thus assume, for illustration purposes, that the binary would have a mass ratio of $\sim$ 0.5. For the system not to fill its Roche lobe---as this would imply a mass transferring system showing signs of accretion, which are not seen---the separation between the two stars should be larger than $\sim$ 3.5~$R_{\odot}$, with a minimal orbital period of the order of 0.5 days. In that case, the maximum orbital period would be 100~km/s. Thus, a star whose radial velocity differs  from that of the cluster  by up to 100 km/s could still be a member, if it were a close binary; in this case, however, we would expect that its radial velocity would change between two epochs. If we now consider the typical, post-mass-transfer, long-period binaries, with periods around 1000 days like those found to constitute the bulk of the BSs in NGC~188, we would expect a maximum radial velocity of 10--13~km/s. In this case, the difference in velocities between two epochs should be very small, i.e., below 1~km/s. Of course, it is possible to have a binary system in between these two cases. These considerations led us to define the following, rather conservative approach to confirm membership of BS stars in Collinder~261:
\begin{itemize}
\item If the individual radial velocities are, given their error bars, compatible with the cluster mean $V_{R}$, and do not change significantly over the four epochs, it is considered a possible single-star member, the outcome of a collision or a merger. It could of course also be a binary with a long period---greater than $\sim 1000$ days. These stars are classified \textbf{``M''}. 
\item On the other hand, if the individual radial velocities are, given their error bars, within 100~km/s with respect the cluster mean $V_{R}$, then:
\begin{enumerate}

\item If the velocities differ more than 20~km/s from $V_{R}$ and  change significantly between two epochs, we can consider the star as a candidate for being a close-binary member of the cluster, \textbf{``M, CB''}. 
\item If the velocities are within 20 km/s from $V_{R}$ and do not change by more than a few km/s between epochs (depending on the possible period, which is constrained by the difference with $V_{R}$), we  possibly have a long-period member (period above 100 days): \textbf{``M, LP''}. 
\end{enumerate}
\item The membership status of the binaries (CB and LP)
can only be secured once we have determined the full orbital solution, and thus derived the systemic velocity.  If none of the above apply, we consider the star to be a non-member, \textbf{``NM''}.
\end{itemize}

\section{RESULTS}\label{sec:results-general}
\subsection{Photometric detections}

We identified 53 possible blue stragglers. Only six of them are in common with the BS population listed by AL07 for Collinder~261. Seven of our stars have already been noted as blue stragglers and binaries in previous studies \citep{Mazur_1995, Vats_2017}, see Table \ref{tab:BS_phot}.

\subsubsection{Radial distribution}
The BS radial distribution has been found to be a powerful tool to estimate the dynamical age of stellar systems (F12 and \citealt{Beccari_2013}). In fact, due to their masses---significantly larger than the average---and their relatively high luminosities, BSs are the ideal objects to measure the effect of dynamical processes, like dynamical friction and mass segregation \citep{Mapelli_2006}. In order to investigate the cluster dynamical state, we studied the BS radial distribution and compared it to those of red giant branch (RGB), red clump (RC), and main sequence (MS) stars, taken as representatives of the normal cluster population, and that therefore  are expected to follow the cluster distribution. 

\begin{figure*}
\includegraphics[width=\linewidth]{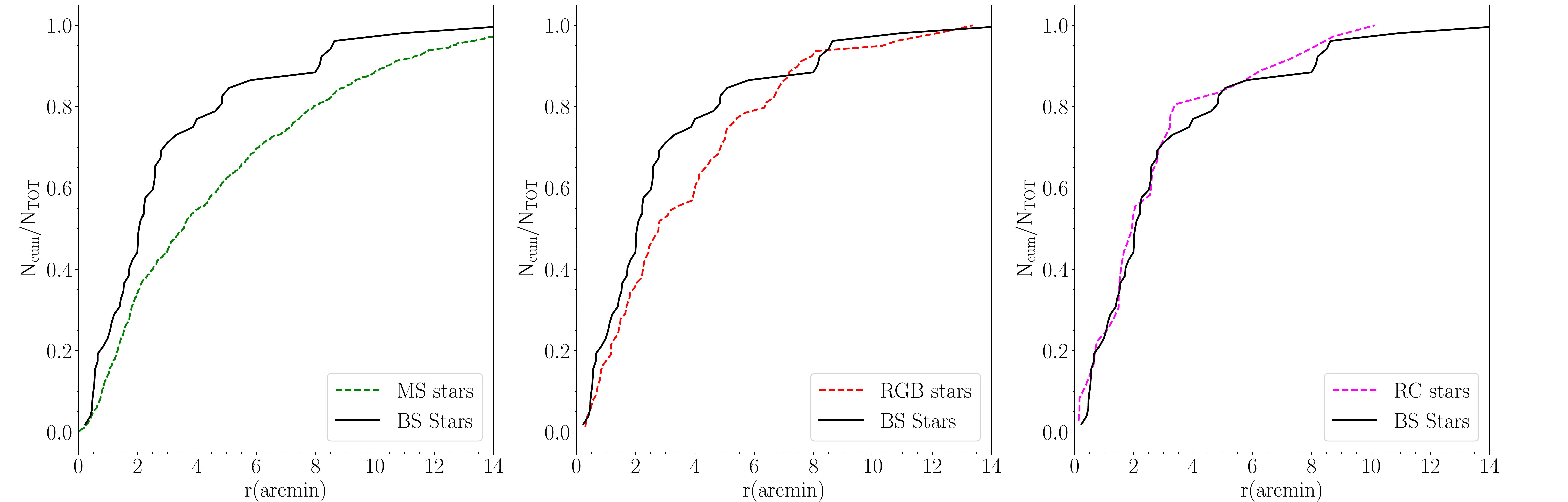}
    \caption{Cumulative spatial distribution of BSs (black solid lined),  main sequence stars (green dashed lined), red giant stars (red dashed line), and clump stars (magenta dashed line) in Collinder 261.}
    \label{fig:spatial_distri}
\end{figure*}

Main sequence  stars were selected from a region free of binary contamination.  We considered as MS stars all those 
 in the range  $17 < G < 18$. For RGB stars, we selected stars lying in the region $G_{\mathrm{TO}}< G < G_{\mathrm{TO}}-2.5$, with $G_{\rm TO} \sim 16$ (see Figure \ref{fig:cmd+propermotions}). This allowed us to obtain a populous sample of 
reference stars in the same range of $G$ magnitude of BSs, i.e., in the same degree of completeness. Red clump stars were selected from the region between $G_{\mathrm{TO}}-2.5 < G < 13$. As we already 
mentioned, the accurate astrometric solutions of \emph{Gaia} let  us  identify the stellar population of Collinder~261 in a very reliable way. We identified 53 BS,  79 RGB, 37 RC, and 833 MS stars. \\

The cumulative spatial distribution of the samples are shown in Figure \ref{fig:spatial_distri},  where the different panels show the normalized cumulative distribution of the BS candidates (black solid line), 
compared to those of MS stars (green dashed line), RGB stars (red dashed line) and RC stars (magenta dashed line). The BSs  appear more centrally concentrated  than the MS and RGB stars.  The higher concentration of BSs that we find in the cluster internal region relative to the evolved stars, has already been observed in other open clusters \citep{Geller_2008, Bhattacharya_2019}. The BS population in Collinder~261 is centrally concentrated, within about 5$'$ with respect to MS an RGB stars. In the right panel of Figure \ref{fig:spatial_distri}, clump stars and BSs have approximately the same distribution. In terms of mass segregation, BSs should be more centrally concentrated, given their combined masses---higher than that of TO stars, than RC stars---slightly less massive than  TO stars. We suggest that the similarity of the radial profiles of BSs and RC stars is possibly  due to the small-number statistics in our sample for both populations. It is worth mentioning that
\citet{Carraro_2014-1} found marginal evidence that the BSs of the old open cluster Melotte~66
 are more concentrated than its clump stars, as in Collinder~261.\\

To quantify whether the radial distributions of BS, RGB, RC, and MS stars are extracted from the same parent distribution, thus indicating an absence of segregation, we used the $k$-sample Anderson-Darling  test \citep[hereafter A-D test]{Scholz_1987}. The A-D test is similar to the more widespread Kolmogorov-Smirnov test, but has a greater sensitivity to the tails of the cumulative distribution. The A-D test indicates a difference of 99.9\% between the distributions of BS and MS stars, and a difference of 86.9\% between the distributions of BS and RGB stars, which therefore favor the  existence of a real mass segregation in Collinder~261. On the contrary, we do not find the population of BS stars to be centrally concentrated with respect to RC stars, as the A-D test gives a probability of 16.0\% that both populations do not originate from the same distribution. The same observation was found by \citet{Carraro_2014-1} in  Melotte~66. \citet{Mazur_1995} discovered 45 short-period eclipsing binaries in the Collinder~261 field, and estimated the frequency of the binary BSs among their sample within~6~arcmin from the cluster center. They found a frequency of between 11\% and 28\%, supporting the hypothesis that a significant fraction of BSs formed as a result of mass transfer in close binary systems. Our analysis also supports this scenario, in which BS stars in Collinder~261 are primordial binaries sinking towards the cluster center due to their combined mass, much larger than that of normal cluster stars.

\begin{figure*}
	\includegraphics[width=2.1\columnwidth]{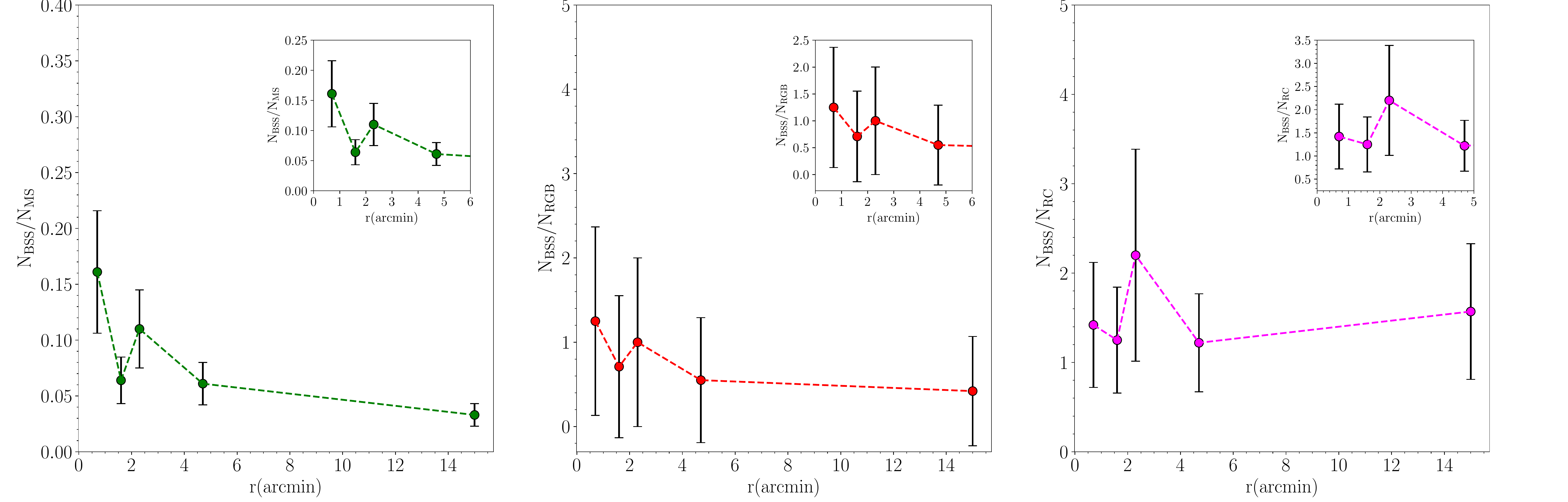}
    \caption{Number of BSs with respect to that of the reference stars: green for main sequence stars, red for red giant stars and magenta for clump stars, plotted as a function of the distance from the cluster center expressed in arcmin. Errors are Poisson.}
    \label{fig:specific_freq}
\end{figure*}

A further indicator of segregation is the number of BSs normalized to the number of a reference population.  We divided the field of view in 5 concentric annuli, each one containing  approximately the same number of BSs  ($\sim 11$). Star counts are in Table \ref{tab:ratioBS_X}. Figure \ref{fig:specific_freq} plots the number of BS candidates with respect to that of reference stars in each annulus, as a function of the  radial distance in arcmin, and the colors are the same described above. When we compare the RC population and BS stars, we note  a maximum at $r \sim 3'$ (see right panel of Figure \ref{fig:specific_freq}); however, when we consider the errors (Poisson errors) this peak disappears and the distribution becomes flat. The same behavior was observed when comparing BS with MS and RGB populations. Both ratios show a maximum value in the annuli closer to the center, followed by a minimum. The ratios  within $6'$  display two minima at $r \sim 2'$ and $r \sim 5'$,  and two maximum peaks at  $r \sim 1'$ and $r \sim 3'$. As in the previous case, these distributions become flat considering the errors. A statistical test needs to be performed in the future,  to evaluate the degree of BSs segregation with respect to the reference populations.

\begin{table*}
\centering
\label{tab:ratioBS_X}
 \caption{Sub-populations star counts.}
 \begin{tabular}{c c  c  c  c  c c c} 
 \hline 
 Range ($''$) & $N_{\rm BS}$ & $N_{\rm MS}$ & $N_{\rm RC}$ & $N_{\rm RGB}$ & $N_{\rm BS}$/$N_{\rm MS}$ & $N_{\rm BS}$/$N_{\rm RC}$ & $N_{\rm BS}$/$N_{\rm RGB}$ \\
  \hline
 0--42     & 10   & 62    &  7    & 8     & 0.161   &  1.42  & 1.25\\
 42--96      & 10   & 156   &  8    & 14    & 0.064   &  1.25  & 0.71\\
 96--138     & 11   & 100   &  5    & 11    & 0.110   &  2.20  & 1.00\\
 138--282    & 11   & 179   &  9    & 20    & 0.061   &  1.22  & 0.55\\
 282--900    & 11   & 326   &  7    & 26    & 0.033   &  1.57  & 0.42\\
 \hline
 \end{tabular}
 \end{table*}
 
A cluster made up of stars of the same mass is dynamically relaxed on a timescale of $t_{\mathrm{relax}} \sim t_{\mathrm{cross}}\, N/6\log(N)$, where $t_{cross} \sim D/\sigma_{v}$ is the crossing time, $N$ is the total number of stars, and $\sigma_{v}$ is the velocity dispersion \citep{Binney_1987}. The time $t_{\mathrm{relax}}$ is the characteristic scale  in which the cluster reaches some level of kinetic energy equipartition with massive stars sinking to the core, and low-mass stars being transferred to the halo, so mass segregation in a star cluster scales with the relaxation time. Using the standard deviations of the projected proper motions ($\sigma_{\mu_{\alpha}} = 0.13$~mas/yr and $\sigma_{\mu_{\alpha}\cos(\delta)} \simeq 0.14$~mas/yr), and the sizes of the cluster core and half-mass radius reported in Table~\ref{tab:Cr261_parameters}, we obtain, for the cluster core, $t_{\mathrm{relax,c}} \sim 100$~Myr, and for the half-mass radius $t_{\mathrm{relax,h}} \sim 250$~Myr. These values are significantly smaller than the estimated age of Collinder~261 ($\sim$ 6--7 Gyr). Consequently, this could explain the presence of mass segregation in this cluster, particularly in the core.

We made and attempt to link our findings for Collinder~261 with those of F12 for globular clusters. We could not classify this cluster into any of the three families defined by F12 (Section~\ref{sec:intro}), given  the small number of BSs we have. \citet{Ferraro_2012} discovered a tight anti-correlation between the core relaxation time and the shape of the radial distribution. Clusters with a flat distribution---i.e., that show no signs of segregation of its BSs---should have a relaxation time of the order of the age of the Universe. Our findings  do not match with their results, given that the small relaxation time we derive for  the core of Collinder~261 is not compatible with a flat distribution.

\begin{figure}
\centering 
\includegraphics[width=1.1\linewidth]{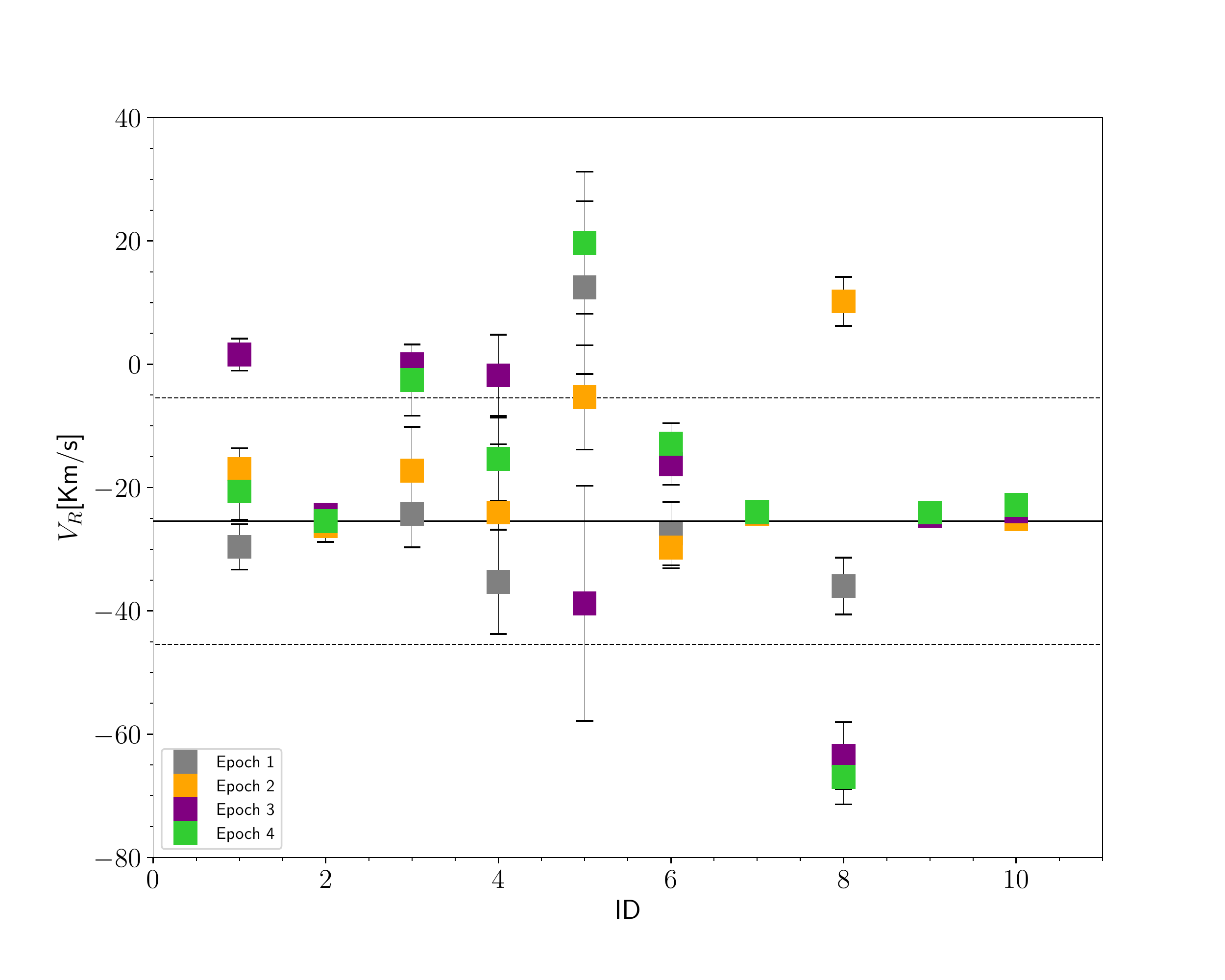}
\caption{Radial velocity distribution of BS candidates in Collinder~261. Different colors indicate different epochs. The solid line represents the cluster mean radial velocity, while the two dashed lines indicate the $\pm$ 20 km/s band we discuss in the text. Errors plotted are the values given by \texttt{fxcor}.}
    \label{fig:radialvelocity-distribution}
\end{figure}

\subsection{Spectroscopic detections}

There are 10 stars in common between those observed with FLAMES and those in our proper motion and parallax selection. Out of our 53 BS candidates, only nine were observed with FLAMES. One YS candidate was also observed, and we decided to study its variability as well. Based on their radial velocity variations, we attempted to roughly  assess their binary nature, namely, to decide if they may be close  or long-period binaries. Our results are plotted in Figure \ref{fig:radialvelocity-distribution}. All the probable binaries would need additional spectroscopic follow-up to be properly characterized, given the small number of observations. In total we found one long-period system, five close-binary systems, and three BS and one YS candidates without variations in  their radial velocity measurements.

\subsubsection{Long-period binaries}

Previous studies have revealed that the BS population in open clusters mostly contains  long-period binaries \citep{Geller_2009}. This kind of stars have periods ranging from a few days to  decades, or even centuries, and it is very difficult to detect them  spectroscopically and photometrically. \\ 

Star BS6 (\%100) is a member according to our astrometric criteria and, according with Figure \ref{fig:radialvelocity-distribution}, is a possible long-period binary stars (M, LP), is the only M, LP we found among our sample. This star is at 2.55$'$ from the cluster center. 

\subsubsection{Close binaries}\label{subsec:CB}

Stars BS1, BS3, BS4, BS5 and BS8 are cluster members and are also classified as possible close-binary systems (M, CB). The presence of these stars within the BS populations is quite unknown; according with previous studies perform in OCs,  they are less numerous than long-period binaries (e.g., NGC~188 \citep{Mathieu_2015}, M67 \citep{Latham_2007}), and their evolutionary histories involve dynamical encounters. According to \citet{Mazur_1995}, Collinder~261 is the star cluster that possesses the richest and  most diversified population of short-period binaries found so far.

Stars BS1 (90\%) and BS3 (\%100) are at  2.50$'$, 2.58$'$ respectively from the cluster center.

Star BS4 (90\%)  has already been classified as an eclipsing binary of Algol type (i.e., detached, cf.~Table~\ref{tab:BS_phot}). The ASAS-SN Variable Stars Database\footnote{\url{https://asas-sn.osu.edu/variables}} \citep{Jayasinghe_2019} gives an amplitude value of 1.87~mag. \citet{Mazur_1995} give a period value of $P\sim 0.49135$ days. This star is at 0.53$'$ from the cluster center.

Star BS5 (100\%) is a very well studied member of Collinder~261. It has already been 
classified as a semi-detached, eclipsing binary of $\beta$~Lyrae type \citep{Avvakumova_2013}. According to  \citet{Samus_2017} and \citet{Jayasinghe_2019}, this binary has a period $P$ of $\sim 0.8040604$ days and a amplitude of 0.44~mag. This star lies at 1.70$'$ from the  cluster center and, according to its upper limit of $v \sin(i)$,  it is a fast rotator (see Table \ref{tab:rv_blue})

Star BS8 (100\%) is a completely different case. In the literature it has been cataloged 
as a detached, eclipsing binary of Algol type \citep{Mazur_1995, Avvakumova_2013, Samus_2017}, and also as a genuine BSS according to AL07.
Most Algol variables are quite close binaries, and therefore their periods are short, of typically 
a few days. Is also very well known that these stars are among the most active and X-ray luminous. \citet{Vats_2017} give an X-ray luminosity $L_{X,u}$ $\sim$ 8.05 (unidades), and \citet{Mazur_1995} a 
preliminary value of the period $P\sim 2.11$~days. This star lies at 3.29$'$ from the cluster
center.\\


\subsubsection{Non-radial velocity variables BSS}

There are three blue stragglers in Collinder~261 that do not show significant radial-velocity variability: BS2 (100\%), BS7 (100\%) and BS9 (100\%). They lie at 1.43$'$, 0.49$'$ and 0.46$'$ respectively from the cluster center. The upper limit we obtained for their rotational velocities are reported in Table \ref{tab:rv_blue}. Is possible that these non-velocity-variable blue stragglers are indeed long-period binaries, perhaps outside of our detection limit. Star BS2 is the bluest BS in our sample and BS7 is the reddest. \\


\subsubsection{Yellow Straggler}

Star YS1 (100\%) is considered to be a yellow straggler according to its position in the color-magnitude diagram, and because it lies beyond the red limit $(G_{\rm BP}-G_{\rm RP}) \sim 1.2$ for the BSs and red giants (see Section \ref{subssec:identification_BSS}). This star is located at 1.67$'$ from the center.  

\subsection{Mass estimations} 

We performed a mass estimation for BS4, BS5, and BS8. These stars are identified in the literature as eclipsing binaries, and are briefly described in Subsection \ref{subssec:identification_BSS}. To fit the orbits we used the radial velocities  we obtained in Subsection \ref{subsec:rv} and the periods reported  in the literature (Subsection \ref{subsec:CB}). For BS5 we divided  the period by 2 (i.e., $P\sim 0.402$~days). For all  stars we assumed that they are members, and that the mean velocity of the cluster as the  systemic velocity $\gamma$ of the binary. \\

\begin{figure}
	\includegraphics[width=\columnwidth]{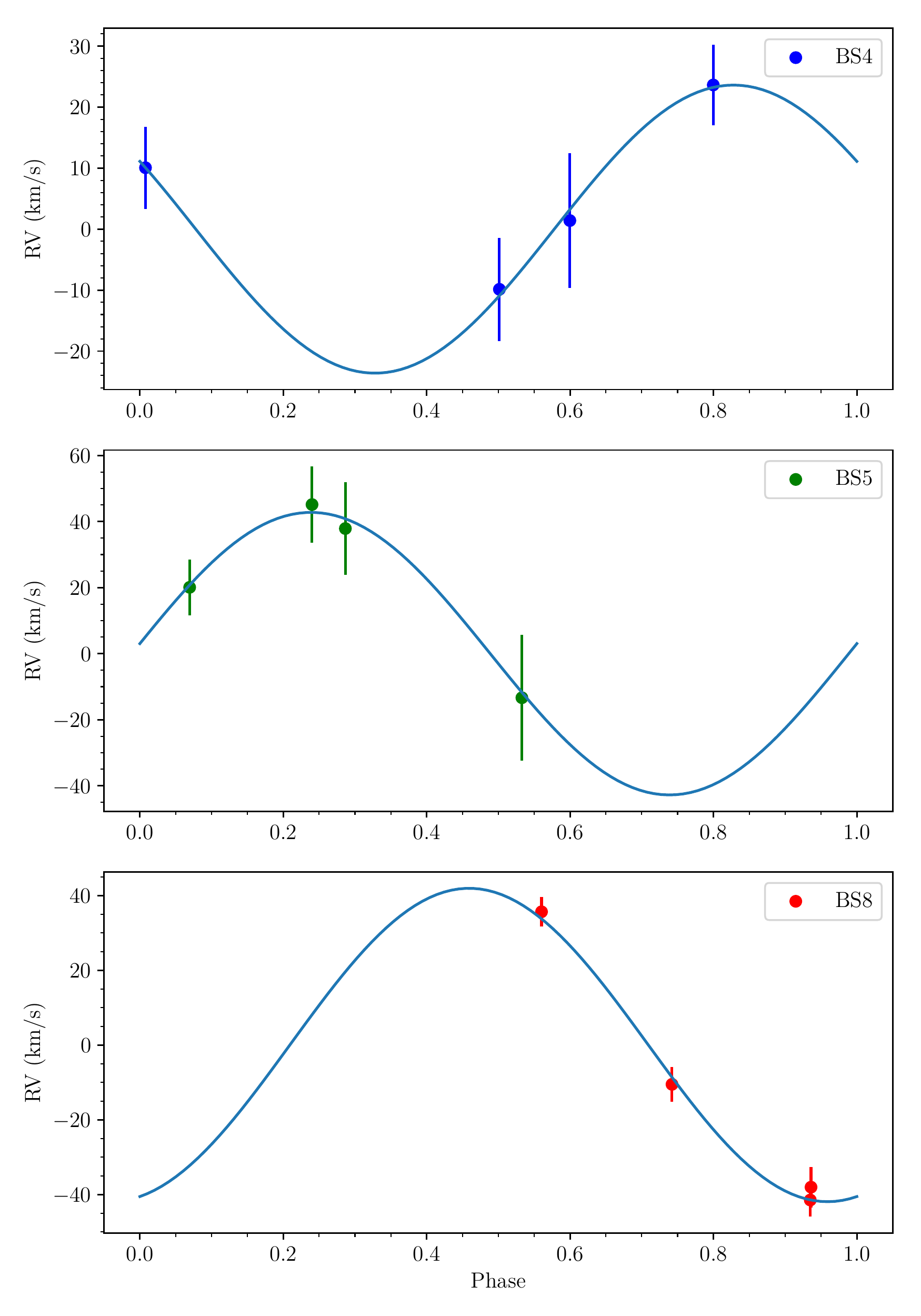}
    \caption{Radial velocity curves of BS4, BS5 and BS8. }
    \label{fig:binary}
\end{figure}

To obtain the mass of the secondary ($M_{2}$) stars, we first  estimated the masses of the primaries ($M_{1}$) from the color-magnitude diagram. The masses we finally derived for BS4 are $M_{1}$ $\sim$ 1.5 M$_{\odot}$ and M$_{2}$= 0.118 $\pm$ 0.005 M$_{\odot}$, for BS5 are $M_{1}$ $\sim$ 1.51 M$_{\odot}$ and M$_{2}$= 0.21 $\pm$ 0.01 M$_{\odot}$, and for BS8 are $M_{1}$ $\sim$ 1.51 M$_{\odot}$ and M$_{2}$= 0.42 $\pm$ 0.02 M$_{\odot}$. Figure \ref{fig:binary} shows the radial velocity curve for each star.

\section{SUMMARY AND CONCLUSIONS}\label{sec:conclusions}

With the accurate data available from the \emph{Gaia} DR2  we have studied the blue straggler  population of the old open cluster Collinder~261. We found 53 blue stragglers and three yellow stragglers among the cluster members. Using red clump stars we calculated the mean proper motions and parallax values, and obtained $\mu_{\alpha}\cos\delta = -6.35 \pm 0.13$ mas/yr, $\mu_{\alpha} = -2.73 \pm 0.14$ mas/yr, and $\varpi = 0.321\pm0.019$ mas, in agreement with the literature. Our results are shown in Figure \ref{fig:cmd+propermotions} and Table \ref{tab:BS_phot}.
 
Following \citet{Bhattacharya_2019}, we used our candidates as test-particles to probe the dynamical state of Collinder~261. In particular, we found that BSs are more centrally concentrated than RGB and MS stars (see Figure \ref{fig:spatial_distri}), and that they follow almost the same distribution than RC stars. To search for mass segregation, we normalized the BS population to several  reference populations (RGB, MS, and RC stars, see Table \ref{tab:ratioBS_X}); before each comparison, we performed an A-D test to check that the populations are not extracted from the same parent distribution, which  should also indicate the presence of segregation. The test gives 99.9\%, 86.9\% and 16.0\%  of MS, RGB and, RC respectively are not drawn from the same distribution than BSs. We found pronounced peaks and minima in the central regions of the cluster, similar to the ones found by \citet{Bhattacharya_2019} in Berkeley~7. However, when taking into account the errors involved,  these results were dismissed. We calculated the relaxation time of our cluster and  found $t_{\rm relax_c} \sim 100$ Myr for the core, and $t_{\rm relax_h} \sim 250$~Myr for the half-light radius. Both values are quite small compared  with the evolutionary age of our cluster (6--7~Gyr). Given these results, we were not able to place Collinder~261 in any of the families defined for globular clusters by \citet{Ferraro_2012}. 

 On the second part of the paper, we have presented the first high-resolution spectroscopic study of the BS population of Collinder~261, adopting membership criteria more solid than the simple photometric ones. 
 So far, spectroscopic studies have been  limited to a very small number of clusters. This is mostly because open clusters are believed to harbor many BS stars \citep{Ahumada_2007}, and therefore studying them in detail represents a huge observational challenge. The reason, however, for which they host so many BS stars, has not been fully deciphered yet. For this study we obtained four epochs of radial velocities;  based on their variations, we separated these stars into candidate members, probable close binaries, and long-period binaries. All the probable binaries would need additional spectroscopic follow-up to be properly characterized, given the small number of epochs available. Unfortunately, the data that we have just presented cover only nine stars of the 53 possible BSs found in our analysis with \emph{Gaia,} and one star  of the three YS candidates we identified. This is so because we originally used the photometric-based list of   \citet{Ahumada_2007} 
 to select the objects to be observed with FLAMES,  a list that differs significantly  from that defined in this work.

Collinder~261 is a cluster of intermediate richness (see Figure \ref{fig:cmd+propermotions}). Our spectroscopic results are shown in Figure \ref{fig:radialvelocity-distribution} and Table \ref{tab:rv_blue}. Radial velocities for four epochs are available for 10 stars, among which we identified five as probable CB, and only one as LP binaries. Four stars did not show radial velocity variations, among them one yellow straggler and three blue stragglers. We performed  an estimation of the masses of BS4, BS5 and BS8 we obtain 0.118 $\pm$ 0.005 M$_{\odot}$, 0.21 $\pm$ 0.01 M$_{\odot}$ and M$_{2}$= 0.42 $\pm$ 0.02 M$_{\odot}$ respectively. We strongly suggest that at least one attempt to fit an orbit solution for the others BSs be made, since it will definitely help understand better the formation history and survival channels of blue straggler stars.  \\

\section{Acknowledgements}
We are grateful to the anonymous referee for helpful comments which significantly helped improve the paper.\\
MJ.~R is supported by CONICY PFCHA through Programa de Becas de Doctorado en el extranjero- Becas Chile / \texttt{2018-72190617}.
\\S.~V gratefully acknowledges the support provide by Fondecyt reg. \texttt{n. 1170518.} \\
J.~A. wishes to thank ESO for stays in the Santiago Headquarters in January 2010, November 2011, and October 2015, where part of this work was originally developed.

\software{IRAF (Tody 1986, Tody 1993), UPMASK (Krone-Martins \& Moitinho 2014), Topcat (Taylor 2005), SPECTRUM (Gray \& Corbally 1994), ATLAS9 (Castelli \& Kurucz 2003)}

\bibliography{sample63}{}
\bibliographystyle{aasjournal}



\end{document}